\documentclass[12pt,draftcls,onecolumn]{IEEEtran}
\ifCLASSINFOpdf
\else
\fi

\usepackage{epsfig}

\usepackage{colortbl}
\usepackage{mathrsfs}
\usepackage{amssymb}
\usepackage{amsmath}
\usepackage{amsfonts}
\newtheorem{theorem}{Theorem}

\newtheorem{definition}{Definition}
\newtheorem{lemma}{Lemma}
\newtheorem{corollary}{Corollary}

\newtheorem{remark}{Remark}
\newtheorem{example}{Example}

\begin{document}

\title{Delta-operator based consensus analysis of multi-agent networks with link failures}


\author{Xue Lin, Yuanshi Zheng, and~ Long Wang

\thanks{This work was supported by NSFC
(Grant Nos. 61375120 and 61533001), the Fundamental Research Funds for the Central Universities
(Grant No. JB160419) and the Young Talent Fund of University Association for Science and Technology in Shaanxi of China (Grant No. 20160208).
(Corresponding author: Long Wang.)}

\thanks{X. Lin and Y. Zheng are with the Center for Complex Systems, School of Mechano-electronic Engineering, Xidian University, Xi'an 710071, China (e-mail:xuelinxd@yeah.net; zhengyuanshi2005@163.com)}
\thanks{L. Wang is with Center for Systems and Control, College of Engineering, Peking University, Beijing 100871, China (e-mail:longwang@pku.edu.cn)}}

\maketitle

\begin{abstract}
In this paper, a discrete-time multi-agent system is presented which is formulated in terms of the delta operator.
The proposed multi-agent system can unify discrete-time and continuous-time multi-agent systems.
In a multi-agent network, in practice, the communication among agents is acted upon by various factors.
The communication network among faulty agents may cause link failures, which is modeled by randomly switching graphs.
First, we show that the delta representation of discrete-time multi-agent system
reaches consensus in mean (in probability and almost surely) if the expected graph is strongly connected.
The results induce that the continuous-time multi-agent system with random networks can also reach consensus in the same sense.
Second, the influence of faulty agents on consensus value is quantified under original network.
By using matrix perturbation theory, the error bound is also presented in this paper. Finally, a simulation example is provided to
demonstrate the effectiveness of our theoretical results.

\end{abstract}


%

\IEEEpeerreviewmaketitle

\section{Introduction}
\IEEEPARstart{D}{istributed} cooperative control problem of multi-agent systems has captured great attention.
This interest is motivated by its diverse applications in various fields, from biology
and sociology to control engineering and computer science. In order to finish different cooperative tasks,
a variety of protocols have been established for multi-agent systems \cite{IEEE-T-AC-2004-1520--1533,IJC-2015-1--13,IEEE-T-AC-2011-1226-1231,Lin--2016--Finite}.
Lots of criteria concerning multi-agent coordination have been provided \cite{IEEE-TAC-2010-950--955,IJC-2014-1--8,A-2009-2605--2611}, etc.

As a fundamental problem of multi-agent coordination, consensus characterizes a phenomenon that
multiple agents achieve a common decision or agreement.
For consensus problem, it has been studied for a long time in management science \cite{JASA--1974-118-121}.
The rise of consensus problem in control filed is influenced by Vicsek model \cite{PRL-1995-1226-1229},
which is a discrete-time model of multiple agents and each agent updates its state by using average of its own state as well as its neighbors'.
The theoretical analysis of consensus for Vicsek model was finished in \cite{IEEE-T-AC-2003-988-1001}.
And then in \cite{IEEE-T-AC-2004-1520--1533}, the authors proposed classical consensus protocols for multi-agent systems
and provided several sufficient conditions to solve the consensus problem.
Inspired by these results, many researchers devoted themselves to studying consensus problems \cite{IEEE-T-AC-2005-655--661,IEEE-T-AC-2008-1804--1816}.
For a multi-agent system, it can be analyzed from two perspectives: one is dynamic model and the other is interaction network.
From the viewpoint of dynamic model, the related researches include first-order dynamics \cite{IEEE-T-AC-2004-1520--1533,IEEE-T-AC-2005-655--661}, second-order dynamics \cite{IJRNC-2007-941--959}, hybrid dynamics \cite{zheng--2017--hybrid}, switched dynamics \cite{AR-2016-314--318},
heterogeneous dynamics \cite{IET-2011-1881--1888}, etc.
From the viewpoint of interaction network, the related researches have fixed networks \cite{IET-2011-1881--1888},
switching networks \cite{IEEE-T-AC-2008-1804--1816}, antagonistic networks \cite{AC-2013-935-946}, random networks \cite{IEEE-TAC-2005-Hatano-1867--1872,IJC-2016-LIN}, and so on.

With the development of digital controller, in many cases, a continuous-time multi-agent system only obtains input signal
at the discrete sampling instants. According to actual factor, researchers investigated the sampled control and
the event-triggered control for multi-agent systems \cite{IEEE-T-AC-2011-1226-1231,ACC-2009-3902--3907,AC-2015-2452--2457}.
It is well known that some discrete-time multi-agent systems are obtained directly from continuous-time multi-agent systems
based on sampled control, which are described by the shift operator.
However, some applications may possess higher sampling rate, which will lead to ill-conditioning problems
when the shift operator is applied to represent the discrete-time system \cite{Proceeding-1992-240-259}.
And the shift operator can't show the intuitive connection between the discrete-time system and the continuous-time system.
To overcome these limitations, Goodwin et al. used the delta operator to represent the dynamics of sampled data system \cite{Proceeding-1992-240-259,Goodwin-1996}.
Compared with shift operator approach, delta operator has several advantages \cite{Proceeding-1992-240-259,Goodwin-1996,AC-1986-1015-1021}, such as superior finite world length coefficient representation and convergence
to its continuous-time counterpart as the sampling period tends to zero.
It is worth pointing out that the delta operator makes the smooth transition from the discrete-time representation to the underlying continuous-time system as sampled period tends to zero. Therefore, it can be used to unify discrete-time and continuous-time systems.
Due to these advantages of the delta operator, there have existed many related research results \cite{AC-2013-2979-2984,IEEE-TIE-2015--5817-5828}. Inspired by these researches, we apply the delta
operator to describe the multi-agent system with sampled data.
A discrete-time representation is proposed for multi-agent systems.

It is well known that the communication may be destroyed in realistic multi-agent network due to
link failures, node failures, etc. Thus, the consensus of multi-agent systems with random networks was  also studied in
\cite{IEEE-TAC-2005-Hatano-1867--1872,IJC-2016-LIN,IEEE-SP-2008-Kar-3315--3326,IEEE--2008--634-649}.
Based on the delta operator, we consider the consensus of multi-agent systems with random networks in this paper.
We assume that there exist faulty agents that only receive information or send information,
which lead to link failures of the network. The original network without faulty agents is an undirected connected graph.
This phenomenon often occurs in practice. For instance, the receiver (emitter) of the agent is failure, which leads to the link failure of the communication network.
Different from \cite{IJC-2016-LIN}, we consider the consensus of discrete-time multi-agent system with directed random networks.
Due to the variation of networks, however, the consensus value is changed.
Therefore, we analyze the influence of faulty agents on the original network.
The main contribution of this paper is twofold.
First, we show that the delta representation of discrete-time multi-agent system reaches
consensus in different sense (in mean, in probability and almost surely) if the expected graph is strongly connected.
Based on the delta operator, we get that the consensus conditions are also appropriate for the
continuous-time multi-agent system with random networks. Second, we analyze the influence of faulty agents on the consensus value under original network.
By using matrix perturbation theory, the error bound between consensus values under
network with link failures and original network is presented.

The structure of this paper is given as follows. In Section 2, based on the delta operator,
a discrete-time multi-agent system is established.
In Section 3, consensus in different sense is studied. In Section 4, we provide the error bound caused by faulty agents.
In Section 5, a simulation example is presented.
Finally, we give a short conclusion in Section 6.

\textbf{Notation:}
Let $\mathbf{1}$, $\mathbf{0}$, $R$ and $R^{n\times n}$ denote the column vector of all ones, the column vector of all zeros,
the set of real numbers
and the $n\times n$ real matrices, respectively.
The $i$th eigenvalue of matrix $A$ can be denoted as
$\lambda_{i}(A)$.
$\|\cdot\|_2$ denotes the standard Euclidean norm, i.e., $\|x(t)\|_2=\sqrt{x^{T}(t)x(t)}$. We write $\|x(t)\|=x^{T}(t)x(t)$.
For the vector 2-norm $\|\cdot\|_2$, the induced matrix norm is $\|A\|_2=\max\limits_{\|x\|\neq0}\frac{\|Ax\|_2}{\|x\|_2}$
$(\|A\|_2=\sqrt{\lambda_{\max}(A^TA)})$.
We write $\|A\|=\lambda_{\max}(A^TA)$.
$B=[b_{ij}]\in R^{n\times n}$, $B\geq0$ if all $b_{ij}\geq0$. We say that $B$ is a
nonnegative matrix if $B\geq0$.
Moreover, if all its row sums are $1$, B is said to be a row stochastic matrix.
For a given vector or matrix $A$, $A^{T}$ denotes
its transpose.
Let $\bar{d}=\max\limits_{i\in\mathcal{I}_n}\{d_{ii}\}$, $H(t_k)=\max\limits_{i\in\mathcal{I}_n}\{x_{i}(t_k)\}$, $h(t_k)=\min\limits_{i\in\mathcal{I}_n}\{x_i(t_k)\}$, $\bar{\lambda}(A)=\max\{\lambda_{i}^{2}(A)\}$ and $\zeta=\max\{\zeta_{t_k}\}$.
$A^{\sharp}$ denotes the group inverse of matrix $A$ \cite{Siam--1975-443-464}.
\section{Preliminaries}\label{problem statement}
\subsection{Graph theory}\label{problem statement 1}
The communication relationship between agents is modeled as a graph
$G=(V, E, A)$ with vertex set $V=\{\nu_1, \nu_2, \ldots, \nu_n\}$, edge set $E=\{e_{ij}\}\subseteq V\times V$
and nonnegative matrix $A=[a_{ij}]_{n\times n}$. If $(\nu_j, \nu_i)\in \textit{E}_i$,
agents $i$ and $j$ are adjacent and $a_{ij}=1$. The set of neighbors of agent $i$
is denoted by $N_i=\{\nu_j|(\nu_j,\nu_i)\}\in E$. The degree matrix $\textit{D}=[d_{ij}]_{n\times n}$
is a diagonal matrix with $d_{ii}=\sum\limits_{j\in N_{i}}a_{ij}$. The Laplacian matrix
of the graph is defined as $L=[l_{ij}]_{n\times n}=D-A$ with $l_{ii}=-\sum\limits_{j\in N_i} a_{ij}$ and $l_{ij}=-a_{ij}$.
The eigenvalues of $L$ can be denoted as $0=\lambda_1(L)\leq\lambda_2(L)\leq\cdots\leq\lambda_n(L)$.
Graph $G$ is said to be strongly connected if there exists a path between any two distinct vertices.
A path that connects $v_i$ and $v_j$ in the directed graph $G$ is a sequence
of distinct vertices $v_{i_0}$, $v_{i_1}$, $v_{i_2}$, $\ldots$, $v_{i_m}$, where $v_{i_0}=v_i$, $v_{i_m}=v_j$
and $(v_{i_r},v_{i_{r+1}})\in E$, $0\leq r\leq m-1$.
When $G$ is an undirected connected graph, then $L$ is positive semi-definite and
has a simple zero eigenvalue. Moreover, there exists $\min\limits_{\xi\neq0, \textbf{1}^{T}\xi=0}\frac{\xi^{T}L\xi}{\xi^{T}\xi}=\lambda_2(L)$ for any $\xi\in R^{n}$.
Throughout this paper, we always assume that $G$ is a undirected connected graph if there does not exist the faulty agent
(agent not be able to receive or send information).

\subsection{Problem statement}\label{problem statement-2016-8-24}

In this paper, we consider a multi-agent system which consists of $n$ agents. The continuous-time dynamics of the $i$th agent is described by
\begin{equation}\label{problem statement eq:1}
\dot{x}_i(t)=u_i(t),\ \  i=1,\ldots,n,
\end{equation}
where $x_i(t)\in R$ and $u_i(t)\in R$ are the state and control input of
$i$th agent, respectively.

For continuous-time multi-agent system (\ref{problem statement eq:1}), a discrete-time representation can be
obtained by using a traditional shift operator. It is given by
\begin{equation}\label{problem statement eq:3}
x_i(t_k+h)=x_i(t_k)+hu_i(t_k),\  \  i=1,\ldots,n,
\end{equation}
where $h$ is the sampling period.
It is worth noting that, as sampling period $h\rightarrow 0$, we lose all information
about the underlying continuous-time multi-agent system (\ref{problem statement eq:1}) \cite{Goodwin--2010--ACC}. Moreover,
it is difficult to describe the next value of $x_i(t_k)$. This difficulty can be
avoided using the delta operator introduced in \cite{Proceeding-1992-240-259}.

%

The delta operator is defined as follows:
$$\delta x(t)=\begin{cases}
\dot{x}(t), \ \ h=0,\\
\displaystyle\frac{x(t+h)-x(t)}{h}, \ \ h\neq0.
\end{cases}$$
Then, by using the delta operator, the discrete-time representation of system (\ref{problem statement eq:1}) is
described by
\begin{equation}\label{problem statement eq:4}
\delta x_{i}(t_k)=u_i(t_k).
\end{equation}
It can be seen that $\delta x_i(t_k)\rightarrow\dot{x}_i(t_k)$ as $h\rightarrow0$. We know that $\delta x_i(t_k)=\dot{x}_i(t_k)$ when $h=0$.
Hence, there is a smooth transition from $\delta x_i(t_k)$ to $\dot{x_i}(t_k)$ as $h\rightarrow0$, which ensures that discrete-time multi-agent system (\ref{problem statement eq:4}) converges to continuous-time multi-agent system (\ref{problem statement eq:1}) as $h\rightarrow0$.

For system (\ref{problem statement eq:1}), we apply the classic consensus protocol $u_i(t)=\sum\limits_{j\in N_i} a_{ij}(x_j(t)-x_i(t))$.
By using zero-order hold, the protocol is given as:
\begin{equation}\label{problem statement  eq:5}
u_i(t_k)=\sum\limits_{j\in N_i} a_{ij}(x_j(t_k)-x_i(t_k)), \ \ t\in[t_k, t_k+h).
\end{equation}

Denote $x(t)=[x_1(t), x_2(t),\ldots,x_n(t)]^{T}$. System $(\ref{problem statement eq:4})$ with protocol (\ref{problem statement  eq:5})
can be represented by
\begin{equation}\label{problem statement eq:6}
\delta x(t_k)=-Lx(t_k).
\end{equation}
Based on above discussion and analysis, we know that discrete-time multi-agent system (\ref{problem statement eq:6}) converges to the continuous-time multi-agent system
\begin{equation}\label{problem statement neweq:7}
\dot{x}(t)=-Lx(t)
\end{equation}
as $h\rightarrow0$.
%

In this paper, the original multi-agent network is undirected connected.
We know that each agent is influenced by the information of its neighbours. However, there may exist the agent
that is unable to receive or send information in the network. We call this type of agent as faulty agent in this paper.
Without loss of generality, we assume that agents $1$ and $2$ are faulty agent, while other agents are normal.
That is, they can always receive and send information in the network.
%
%
Two scenarios are considered.

Scenario I : Four cases are considered:
(1) only agent 1 can't receive information;
(2) only agent 2 can't receive information;
(3) agents 1 and 2 cannot receive information simultaneously;
(4) all agents are normal.
Networks $G_{1}$, $G_{2}$, $G_{3}$ and $G_{4}$ correspond to the four cases (1), (2), (3) and (4), respectively.
We assume that $G_i$ randomly switches among distinct networks $G_i\in\{G_{1}, G_{2}, G_{3}, G_{4}\}$.
Networks $G_{1}$, $G_{2}$, $G_{3}$ and $G_{4}$ correspond to the occurrence probabilities
$1>\alpha>0$, $1>\beta>0$, $1>\gamma>0$ and $1>\theta>0$, respectively. Moreover, $\alpha+\beta+\gamma+\theta=1$.

Scenario II : Four cases are considered:
(1) only agent 1 can't send information;
(2) only agent 2 can't send information;
(3) agents 1 and 2 cannot send information simultaneously;
(4) all agents are normal.
Networks $G^{\prime}_{1}$, $G^{\prime}_{2}$, $G^{\prime}_{3}$ and $G^{\prime}_{4}$ correspond to the four cases (1), (2), (3) and (4), respectively.
We assume that $G^{\prime}_i$ randomly switches among distinct networks $G^{\prime}_i\in\{G^{\prime}_{1}, G^{\prime}_{2}, G^{\prime}_{3}, G^{\prime}_{4}\}$.
Networks $G^{\prime}_{1}$, $G^{\prime}_{2}$, $G^{\prime}_{3}$ and $G^{\prime}_{4}$ correspond to the occurrence probabilities
$1>\alpha>0$, $1>\beta>0$, $1>\gamma>0$ and $1>\theta>0$, respectively. Moreover, $\alpha+\beta+\gamma+\theta=1$.


%
%

System (\ref{problem statement eq:6}) under Scenario I or II can be written as:
\begin{equation}\label{problem statement eq:7}
\delta x(t_k)=
-L_{t_k}x(t_k),
\end{equation}
where $L_{t_k}$ is the Laplacian matrix at time point $t_k$.
Note that the graph $G_{t_k}$ is invariant during the time interval $\Delta$.
Corresponding adjacent matrix at time point $t_k$ is $A_{t_k}$.
Throughout this paper, the sampling period satisfies $\Delta=\bar{k}h$ when $h\nrightarrow0$.

Two main objectives are considered in this paper. First, the consensus of system (\ref{problem statement eq:7}) is considered.
Second, the error bound between consensus values of system (\ref{problem statement eq:6}) and system (\ref{problem statement eq:7}) is presented.
\begin{remark}\label{newlemma1 1}
For simplicity, we focus on two faulty agents. However, the analytical methods concerning error bound
in this paper can be extended to the Scenario of more than two faulty agents, which is left to the interested readers as an exercise.
\end{remark}

\begin{definition}\label{definition 2}System (\ref{problem statement eq:7}) reaches consensus\\
(a) in mean if for any $x_0\in R^{n}$ it holds that
\begin{equation}\label{definition 2 eq:1}
\lim_{t_k \to \infty}E\left[x(t)\right]=v(x_{0});
\end{equation}\\
(b) in probability if $\forall\varepsilon>0$ and any $x_0\in R^{n}$ it holds that
\begin{equation}\label{definition 2 eq:2}
\lim_{t_k \to \infty}P\left\{H(t_k)-h(t_k)\geq\varepsilon\right\}=0;
\end{equation}
(c) almost surely if for any $x_0\in R^{n}$ it holds that
\begin{equation}\label{definition 2 eq:3}
P\left\{\lim_{t_k \to \infty}(H(t_k)-h(t_k))=0\right\}=1.
\end{equation}
\end{definition}
\begin{definition}[\cite{Buckley--2002--33-41}]\label{definition 3}
Let $W$ denote the transition matrix of Markov chain. Then the Markov chain is called the regular chain if there exist $k>0$ such that
$W^{k}$ has only positive elements.
\end{definition}
\begin{lemma}[\cite{Siam--1975-443-464}]\label{2016. newlemma3}
If $T$ is the transition matrix of a regular chain, then $A^{\sharp}=\sum\limits_{k=0}^\infty(T^{k}-T^{\infty})$
where $A=I-T$.
\end{lemma}

\begin{lemma}[\cite{Siam-1979-273-283}]\label{2016. newlemma4}
If C and $\tilde{C}$ are ergodic chains with transition matrices $T$ and $\tilde{T}=T-E$ and limiting probability vectors $s$ and $\tilde{s}$, respectively, then
$s-\tilde{s}=sEA^{\sharp}(I+EA^{\sharp})^{-1}$ where $E\textbf{1}=\textbf{0}$ and $A=I-T$.
\end{lemma}
\begin{lemma}[\cite{Goodwin-1996}]\label{2016. newlemma1}
The property of delta operator for any time function $x(t_k)$
and $y(t_k)$ can be represented as
$$\delta(x(t_k)y(t_k))=\delta (x(t_k))y(t_k)+x(t_k)\delta (y(t_k))+h\delta (x(t_k))\delta (y(t_k)).$$
\end{lemma}

\begin{lemma}\label{2016. lemma1}
Assume that the sampling period $0< h<\frac{1}{d_{max}}$. Then, system (\ref{problem statement eq:6}) can reach average consensus
if the graph is undirected connected.
\end{lemma}
{\it Proof.} Let $\nu(t_{k})=x(t_k)-\frac{\textbf{1}\textbf{1}^{T}}{n}x(0)$. Due to $L\textbf{1}=0$, one has $\delta (\nu(t_{k}))=-L\nu(t_{k})$.
Consider $V(t_k)=\|\nu(t_{k})\|$ as a Lyapunov function.
By Lemma \ref{2016. newlemma1}, it holds that
$$
\begin{array}{rl}
\delta V(t_k)&=\delta^{T}(\nu(t_{k}))\nu(t_{k})
+\nu^{T}(t_{k})\delta (\nu(t_{k}))+h\delta^{T}(\nu(t_{k}))\delta(\nu(t_{k}))\\
&=\nu^{T}(t_{k})(-2L+hL^{T}L)\nu(t_{k})\\
&=\nu^{T}(t_{k})\Xi\nu(t_{k}).$$
\end{array}
$$
Since the graph is undirected connected, which implies that the Laplacian matrix $L$ is positive semi-definite. Hence, the eigenvalues of $\Xi$ are repsented by $-2\lambda_i(L)+h\lambda^{2}_i(L)$.
From Gersgorin Disk Theorem, we get $\lambda_i(L)\leq2d_{max}$. Then $-2+h\lambda_i(L)<-2+\frac{1}{d_{max}}2d_{max}\leq0$.
Owing to $\min\limits_{\xi\neq0, \textbf{1}^{T}\xi=0}\frac{\xi^{T}L\xi}{\xi^{T}\xi}=\lambda_2(L)$, then
$$\delta V(t_k)\leq-(2\lambda(L)-h\lambda^{2}(L))\|\nu(t_{k})\|$$
where $2\lambda(L)-h\lambda^{2}(L)=\min\{2\lambda_2(L)-h\lambda^{2}_2(L),2\lambda_n(L)-h\lambda^{2}_n(L)\}$.
Due to $\lambda_2(L)>0$ and $\lambda_n(L)>0$, this proves that $\delta V(t_k)<0$.
Therefore, $\nu(t_k)$ is converge to $0$.
That is, system (\ref{problem statement eq:6}) can achieve average consensus asymptotically.
$\blacksquare$

\begin{remark}\label{lemma1 1}
From Lemma \ref{2016. lemma1},
there exists $$\delta V(t_k)=\displaystyle\frac{V(t_k+h)-V(t_k)}{h}<0,$$
which implies that
$$\lim_{h \to 0}\delta V(t_k)=\lim_{h \to 0}\frac{V(t_k+h)-V(t_k)}{h}=\dot{V}(t)<0.$$
It can be seen that $\delta V(t_k)<0$ can be reduced to the $\dot{V}(t)<0$ as $h\rightarrow0$.
Note that system (\ref{problem statement eq:6}) converges to system (\ref{problem statement neweq:7}) as $h\rightarrow0$.
Consequently, system (\ref{problem statement neweq:7}) reaches average consensus under
undirected connected graph.
\end{remark}
\section{Consensus analysis}\label{main result1}
In this section, it is shown that system (\ref{problem statement eq:6}) reaches
consensus despite the existence of
faulty agents.
Supposed that Scenario I and Scenario II have the same expression pattern for the network.
Hence, the following results can be viewed as the unified conclusions of system (\ref{problem statement eq:6}) under Scenarios I and II.

\begin{theorem}\label{2016 theorem1}
Assume that the sampling period $0<h<\frac{1}{d_{max}}$. Then, system (\ref{problem statement eq:7})
reaches consensus in mean if the expected graph is strongly connected. Furthermore,
\begin{equation}\label{theorem1 newqueation1 }
\begin{array}{rl}
\lim\limits_{t_k \to \infty}E[x(t_k)]=\textbf{1}\bar{\pi}^{T}x(0),
\end{array}
\end{equation}
where
 \[ \textbf{1}\bar{\pi}^{T}=
\begin{cases}
\textbf{1}\bar{\pi}_1^{T}=\lim\limits_{k \to \infty}W_{1}^{k} :  \  \   W_{1}=E[(I-hL_{t_k})^{\bar{k}}], \\
\textbf{1}\bar{\pi}_2^{T}=\lim\limits_{k \to \infty}W_{2}^{k}:   \  \  W_{2}=E[e^{-L_{t_k}\Delta}], \ \ h\rightarrow0,\ \
\end{cases}  \]
vectors $\bar{\pi}_1>0$ and $\bar{\pi}_2>0$ are left eigenvectors of the matrices $ W_{1}$ and $ W_{2}$, respectively, such that
$\bar{\pi}_1^{T}\textbf{1}=1$ and $\bar{\pi}_2^{T}\textbf{1}=1$.
\end{theorem}
{\it Proof.}
As pointed out in \cite{Proceeding-1992-240-259}, the solution to system (\ref{problem statement eq:6}) is $x(t)=(I-hL)^{\frac{t}{h}}x(0)$.
Due to the invariance of graph $G_{t_k}$ during the time
interval $\Delta$, it can be get that $x(t_k+\Delta)=(I-hL_{t_k})^{\frac{\Delta}{h}}x(t_{k})$. Then
\begin{equation}\label{theorem1 queation1 }
\begin{array}{rl}
&\lim\limits_{k \to \infty}E(x(t_k))\\
&=\lim\limits_{k \to \infty}[E((I-hL_{t_k})^{\bar{k}})]^{k}x(0)\\
&=\lim\limits_{k \to \infty}[(I-hL_{1})^{\bar{k}}\alpha+(I-hL_{2})^{\bar{k}}\beta+(I-hL_{3})^{\bar{k}}\gamma+(I-hL_{4})^{\bar{k}}\theta]^{k}x(0)\\
&=\lim\limits_{k \to \infty}W_{1}^{k}x(0).
\end{array}
\end{equation}
According to $0<h<\frac{1}{d_{max}}$, we have $I-hL_{i}=I-hD_{i}+hA_{i}\geq0$ with positive diagonal elements.
Since $\alpha>0$, $\beta>0$, $\gamma>0$ and $\theta>0$, it is immediate that $E((I-hL_{t_k})^{\bar{k}})$ is also
nonnegative matrix with positive diagonal elements.

It follows that
$$E((I-hL_{t_k})^{\bar{k}})\geq \Pi(A_{1}\alpha+A_{2}\beta+A_{3}\gamma+A_{4}\theta),$$
where $\Pi$ is a positive diagonal matrix.
Since the expected graph is strongly connected, matrix $E((I-hL_{t_k})^{\bar{k}})$ is a nonnegative irreducible
with positive diagonal elements.
Moreover, it is easy to verify that $E((I-hL_{t_k})^{\bar{k}})\textbf{1}=\textbf{1}$. Then,
by Ger$\breve{s}$gorin Disc theorem, one has $|\lambda_{i}(E((I-hL_{t_k})^{\bar{k}}))|\leq1$. Hence,
by Perron-Frobenius Theorem \cite{Horn-2012}, it can be deduced that $\rho(E((I-hL_{t_k})^{\bar{k}}))=1$ is an algebraically simple eigenvalue.
Consequently, matrix $W_1$ is a primitive.
By virtue of Theorem 8.5.1 in \cite{Horn-2012}, we obtain that $\lim\limits_{t_k \to \infty}E[x(t_k)]=\textbf{1}\bar{\pi_1}^{T}x(0).$
Hence, system (\ref{problem statement eq:7})
reaches consensus in mean.

Next, we give the consensus value of system (\ref{problem statement eq:7}) as $h\rightarrow0$.
By Proposition 11.1.3 in \cite{Dennis-2009}, it follows that $x(t_k+\Delta)=(I-hL_{t_k})^{\frac{\Delta}{h}}x(t_{k})=e^{-L_{t_k}\Delta}$ as $h\rightarrow0$.
Hence
\begin{equation}\label{theorem1 queation2 }
\begin{array}{rl}
&\lim\limits_{t_k \to \infty}E(x(t_k))\\
&=\lim\limits_{k \to \infty}[E((I-hL_{t_k})^{\bar{k}})]^{k}x(0)\\
&=\lim\limits_{k \to \infty}(e^{-L_{1}\Delta}\alpha+e^{-L_{2}\Delta}\beta+e^{-L_{3}\Delta}\gamma+e^{-L_{4}\Delta}\theta)^{k}x(0)\\
&=\lim\limits_{k \to \infty}W_{2}^{k}x(0).
\end{array}
\end{equation}
Let $L_{t_k}=d_{max}I-\bar{A}_{t_k}$, then $e^{-L_{t_k}\Delta}=e^{-\Delta d_{max} I}e^{\bar{A}_{t_k}}\geq\zeta_{t_k}\bar{A}_{t_k}$ for $\zeta_{t_k}>0$
where $\bar{A}_{t_k}\geq A_{t_k}$. Hence,
$E(e^{-L_{t_k}\Delta})\geq\zeta(\bar{A}_{1}\alpha+\bar{A}_{2}\beta+\bar{A}_{3}\gamma+\bar{A}_{4}\theta)$. That is, matrix $E(e^{-L_{t_k}\Delta})$ is a nonnegative irreducible with positive diagonal elements.
Similar to the previous discussion, we have
$\lim\limits_{t_k \to \infty}E[x(t_k)]=\textbf{1}\bar{\pi_2}^{T}x(0)$ as $h\rightarrow0$.
$\blacksquare$

\begin{theorem}\label{2016 theorem2}
Assume that the sampling period $0<h<\frac{1}{d_{max}}$. Then, system (\ref{problem statement eq:7})
reaches consensus in probability if the expected graph is strongly connected.
\end{theorem}
{\it Proof.}
Since the expected graph is strongly connected, by Theorem \ref{2016 theorem1},
it follows that $\lim\limits_{t_k \to \infty}E(H(t_k)-h(t_k))=0$.
Let $(I-hL_{t_k})^{\bar{k}}=[w_{ij}]_{n\times n}$, we have $x_i({t_{k}+\Delta})=\sum\limits_{j=1}^n w_{ij}x_j(t_{k})$.
Since matrix $(I-hL_{t_k})^{\bar{k}}$ is a row stochastic matrix, we get $H(t_k+\Delta)\leq H(t_{k})$ and $h(t_k+\Delta)\geq h(t_k)$.
It can be verified that the $H(t_k)-h(t_k)$ is nonincreasing.

Let $t_{k+1}=t_{k}+\Delta$, hence $0\leq H(t_{k+1})-h(t_{k+1})\leq H(t_{k})-h(t_{k})$, which yields
\begin{equation}\label{theorem2 queation1 }
E[(H(t_{k+1})-h(t_{k+1}))^{2}]\leq E[H(t_{k})-h(t_{k})](H(0)-h(0)).
\end{equation}
Hence
\begin{equation}\label{theorem2 queation2 }
\lim\limits_{t_k \to \infty}E[(H(t_{k+1})-h(t_{k+1}))^{2}]=0.
\end{equation}
As a result of Chebyshev’s inequality, for any $\varepsilon>0$, it follows that
\begin{equation}\label{theorem2 queation3}
P\left\{H(t_k)-h(t_k)\geq\varepsilon\right\}\leq\dfrac{E[(H(t_k)-h(t_k))^{2}]}{\varepsilon^{2}}.
\end{equation}
Therefore
\begin{equation}\label{theorem2 queation4}
\lim\limits_{t_k \to \infty}P\left\{H(t_k)-h(t_k)\geq\varepsilon\right\}=0.
\end{equation}
It is shown from Theorem \ref{2016 theorem1} that $\lim\limits_{h \to 0}(I-hL_{t_k})^{\frac{\Delta}{h}}=e^{-L_{t_k}\Delta}$.
Matrix $e^{-L_{t_k}\Delta}$ is also a row stochastic matrix. Similar to the above proof,
it can be proved that (\ref{theorem2 queation4}) also holds as $h\rightarrow0$.
$\blacksquare$
\begin{theorem}\label{2016 theorem3}
Assume that the sampling period $0<h<\frac{1}{d_{max}}$. Then, system (\ref{problem statement eq:7})
reaches consensus almost surely if the expected graph is strongly connected.
\end{theorem}
{\it Proof.}
It follows from Theorem \ref{2016 theorem2} that $H(t_k)-h(t_k)\rightarrow0$ in probability.
By Theorem 2.5.3 in \cite{Robert-2000},
there exists a subsequence of $\{H(t_k)-h(t_k)\}$ that converges almost surely to 0.
Hence, for any $\varepsilon>0$, there exists  $t_{l}$ such that for $t_{\bar{l}}\geq t_{l}$, $H(t_{\bar{l}})-h(t_{\bar{l}})<\varepsilon$ almost surely.
Since $\{H(t_k)-h(t_k)\}$ is nonincreasing, it holds that $0\leq H(t_{\bar{l}+1})-h(t_{\bar{l}+1})\leq H(t_{\bar{l}})-h(t_{\bar{l}})<\varepsilon$
almost surely. Therefore, for any $t_k\geq t_{\bar{l}+1}$, there holds $0\leq\{H(t_k)-h(t_k)\}<\varepsilon$ almost surely. This implies that
system (\ref{problem statement eq:7}) reaches consensus almost surely. $\blacksquare$
\begin{remark}\label{Theorem1 remark 1}
As pointed out in Theorem \ref{2016 theorem1}, one has $x(t_k+\Delta)=e^{-L_{t_k}\Delta}x(t_k)$ as $h\rightarrow0$.
Since the network is invariant during time interval $\Delta$,
partial state of system (\ref{problem statement neweq:7}) can be represented
by $x(t_k+\Delta)=e^{-L_{t_k}\Delta}x(t_k)$.
It is shown from Theorem \ref{2016 theorem1} that the sequence $x(t_k)$ achieves consensus in mean.
Then, using $-L\textbf{1}=0$, we can conclude that $x(t)$ achieves consensus in mean. Therefore, system (\ref{problem statement neweq:7}) with random networks reaches consensus in mean if the expected graph is strongly connected.

This indicates that the consensus result of system (\ref{problem statement eq:7}) with random networks reduces to
the consensus result of system (\ref{problem statement neweq:7}) under random networks as $h\rightarrow0$.
Moreover, Theorems \ref{2016 theorem2} and \ref{2016 theorem3} are also appropriate for the continuous-time multi-agent system as $h\rightarrow0$.
\end{remark}

%

\section{Error analysis}\label{main result2}
In this section, we consider the error bound on the consensus value $\lim\limits_{t_k\rightarrow\infty}E(x(t_k))$ and the consensus value under original network $\frac{\textbf{1}\textbf{1}^{T}}{n}x(0)$,
i.e.,
\begin{equation}\label{Error bound queation2}
\begin{array}{rl}
\lim\limits_{t_k\rightarrow\infty}E(x(t_k))-\frac{\textbf{1}\textbf{1}^{T}}{n}x(0)&=\textbf{1}\bar{\pi}^{T}x(0)-\frac{\textbf{1}\textbf{1}^{T}}{n}x(0)\\
&=\textbf{1}(\bar{\pi}^{T}-\frac{\textbf{1}^{T}}{n})x(0).
\end{array}
\end{equation}
To solve this problem, the matrix perturbation theory and the property of finite Markov chains are applied.

On the analysis of the consensus problem, we apply $E(x(t_k+\Delta))=W_1E(x(t_{k}))$.
Suppose that the expected graph is strongly connected and $0<h<\frac{1}{d_{max}}$. Then, Theorem \ref{2016 theorem1} shows that $W_1$ is a row stochastic matrix such that $W_1^{k}>0$. By property of row stochastic matrix, each element $w_{ij}$ of matrix $W_1$ satisfies $0\leq w_{ij}<1$.
Hence, by Definition \ref{definition 3}, $W_1$ can be regarded as the transition matrix of a regular chain.
Moreover, $W_1$ is a transition matrix of ergodic chain.
It is noteworthy that the following analysis results are appropriate for Scenario I and Scenario II.

\begin{theorem}\label{2016 theorem4}
Assume that the sampling period $0<h<\frac{1}{d_{max}}$ and the expected graph is strongly connected. Then
\begin{equation}\label{theorem4 queation1 }
\begin{array}{rl}
\|\bar{\pi}_{1}^{T}-\frac{\textbf{1}^{T}}{n}\|\leq\|D_{1}\|\frac{1}{1-\bar{\lambda}(\bar{W}_{1}-\frac{\textbf{1}\textbf{1}^{T}}{n})},
\end{array}
\end{equation}
and
\begin{equation}\label{theorem4 newqueation1}
\begin{array}{rl}
\|\bar{\pi}_{2}^{T}-\frac{\textbf{1}^{T}}{n}\|\leq\|D_{2}\|\frac{1}{1-\bar{\lambda}(\bar{W}_{2}-\frac{\textbf{1}\textbf{1}^{T}}{n})},\ \ h\rightarrow0,
\end{array}
\end{equation}
where $\bar{W}_1=(I-hL_{4})^{\bar{k}}$, $D_1=\sum\limits_{i=1}^3((I-hL_{i})^{\bar{k}}-(I-hL_{4})^{\bar{k}})p_i$,
$\bar{W}_2=e^{-L_{4}\Delta}$, $D_2=\sum\limits_{i=1}^3(e^{-L_{i}\Delta}-e^{-L_{4}\Delta})p_i$.
$p_1$, $p_2$, $p_3$ correspond to $\alpha$, $\beta$, $\gamma$, respectively.
\end{theorem}

{\it Proof.}
It is pointed out that
$W_1=(I-hL_{1})^{\bar{k}}\alpha+(I-hL_{2})^{\bar{k}}\beta+(I-hL_{3})^{\bar{k}}\gamma+(I-hL_{4})^{\bar{k}}\theta$ can be written as $W_1=\bar{W}_1+D_1$.
From Theorem \ref{2016 theorem1}, we can derive that
$\lim\limits_{k \to \infty}\bar{W}_{1}^{k}=\frac{\textbf{1}\textbf{1}^{T}}{n}$ and $W_1$ is a row stochastic matrix.
Hence, it proves that the row sums of $D_1$ are all equal to $0$. Moreover, it follows from Theorem \ref{2016 theorem1} that
$\lim\limits_{k \to \infty}W_{1}^{k}=\textbf{1}\bar{\pi}_1^{T}$. Let $e=\bar{\pi}_{1}^{T}-\frac{\textbf{1}^{T}}{n}$, we analyze
the error bound of $\|e\|$.

Using Lemmas \ref{2016. newlemma3} and \ref{2016. newlemma4}, it holds that
\begin{equation}\label{theorem4 queation2}
\begin{array}{rl}
e=\frac{\textbf{1}^{T}}{n}D_1F(I-D_1F)^{-1},
\end{array}
\end{equation}
where $F=\sum\limits_{k=0}^\infty(\bar{W}_1^{k}-\frac{\textbf{1}\textbf{1}^{T}}{n})$. By some algebraic manipulations for (\ref{theorem4 queation2}), the following equation holds
\begin{equation}\label{theorem4 queation3}
\begin{array}{rl}
\frac{\textbf{1}^{T}}{n}D_1F=e-\bar{\pi}_{1}^{T}D_1F+\frac{\textbf{1}^{T}}{n}D_1F,
\end{array}
\end{equation}
i.e., $e=\bar{\pi}_{1}^{T}D_1F$. This implies that $\|e\|=\|\bar{\pi}_{1}^{T}D_1F\|\leq\|\bar{\pi}_{1}^{T}\|\|D_1F\|$.
Due to $\bar{\pi}_{1}>0$ and $\bar{\pi}_{1}^{T}\textbf{1}=\textbf{1}$, we get $\|e\|\leq\|D_1F\|$.

It follows from $F=\sum\limits_{k=0}^\infty(\bar{W}_1^{k}-\frac{\textbf{1}\textbf{1}^{T}}{n})$ that $\|e\|\leq\|D_1F\|=\|D_1\sum\limits_{k=0}^\infty\bar{W}_1^{k}\|$. Hence, we analyze $\|D_1\sum\limits_{k=0}^\infty\bar{W}_1^{k}\|$. To solve this problem, we introduce a vector $y(0)$ such that $\|y(0)\|\neq0$ and
$y(k+1)=\bar{W}_1^{k}y(0)$. It is obvious that $\textbf{1}^{T}y(k)=\textbf{1}^{T}y(0)$. Therefore,
\begin{equation}\label{theorem4 queation4}
\begin{array}{rl}
\|D_1\bar{W}_1^{k}y(0)\|\leq&\|D_1\|\|\bar{W}_1^{k}y(0)\|\\
=&\|D_1\|\|\bar{W}_1y(k)\|\\
=&\|D_1\|\|(\bar{W}_1-\frac{\textbf{1}\textbf{1}^{T}}{n})(y(k)-\frac{\textbf{1}\textbf{1}^{T}}{n}y(0))\|\\
=&\|D_1\|\|(\bar{W}_1-\frac{\textbf{1}\textbf{1}^{T}}{n})^{k}\|\|I-\frac{\textbf{1}\textbf{1}^{T}}{n}\|\|y(0)\|.
\end{array}
\end{equation}
Owing to $\bar{W}_1\frac{\textbf{1}\textbf{1}^{T}}{n}=\frac{\textbf{1}\textbf{1}^{T}}{n}\bar{W}_1$, by Theorem 4.5.15 in \cite{Horn-2012},
we have $\lambda_i(\bar{W}_1-\frac{\textbf{1}\textbf{1}^{T}}{n})=(1-h\lambda_i(L_4))^{\bar{k}}-\lambda_i(\frac{\textbf{1}\textbf{1}^{T}}{n})$.
There exists the eigenvector $\textbf{1}$ corresponding to $\lambda_1(\bar{W}_1)=1$ and $\lambda_1(\frac{\textbf{1}\textbf{1}^{T}}{n})=1$,
which implies $\lambda_1(\bar{W}_1-\frac{\textbf{1}\textbf{1}^{T}}{n})=0$
and $\lambda_i(\bar{W}_1-\frac{\textbf{1}\textbf{1}^{T}}{n})=(1-h\lambda_i(L_4))^{\bar{k}}$, $i=2,\cdots, n$.
Moreover, by a similar analysis, we have $\lambda_1(I-\frac{\textbf{1}\textbf{1}^{T}}{n})=0$ and $\lambda_i(I-\frac{\textbf{1}\textbf{1}^{T}}{n})=1$, $i=2,\ldots,n$.
Due to $0<h<\frac{1}{d_{max}}$, it can be deduced that $-1<1-h\lambda_i(L_4)<1$.
We know that matrix $\bar{W}_1-\frac{\textbf{1}\textbf{1}^{T}}{n}$ is symmetric. Consequently,
$0\leq\bar{\lambda}(\bar{W}_1-\frac{\textbf{1}\textbf{1}^{T}}{n})=
\max\{\lambda_{i}((\bar{W}_1-\frac{\textbf{1}\textbf{1}^{T}}{n})^{T}(\bar{W}_1-\frac{\textbf{1}\textbf{1}^{T}}{n}))\}<1$.
It follows that
\begin{equation}\label{theorem4 queation5}
\begin{array}{rl}
\|D_1\bar{W}_1^{k}y(0)\|\leq&\bar{\lambda}^{k}(\bar{W}_1-\frac{\textbf{1}\textbf{1}^{T}}{n})\|D_1\|\|y(0)\|.
\end{array}
\end{equation}
Then
\begin{equation}\label{theorem4 queation6}
\begin{array}{rl}
\|D_1\sum\limits_{k=0}^\infty\bar{W}_1^{k}y(0)\|\leq &\sum\limits_{k=0}^\infty\|D_1\bar{W}_1^{k}y(0)\|\\
\leq&\|D_1\|(1+\bar{\lambda}(\bar{W}_1-\frac{\textbf{1}\textbf{1}^{T}}{n})
+\bar{\lambda}^{2}(\bar{W}_1-\frac{\textbf{1}\textbf{1}^{T}}{n})+\cdots)\|y(0)\|.
\end{array}
\end{equation}
On account of $\|D_1\bar{W}_1\|_2=\max\limits_{\|y(0)\|_2\neq0}\frac{\|D\bar{W}_1^{k}y(0)\|_2}{\|y(0)\|_2}$ and
$\bar{\lambda}(\bar{W}_1-\frac{\textbf{1}\textbf{1}^{T}}{n})<1$, we obtain
$\|e\|\leq\|D_1\|\frac{1}{1-\bar{\lambda}(\bar{W_1}-\frac{\textbf{1}\textbf{1}^{T}}{n})}$.

When $h\rightarrow0$, from Theorem \ref{2016 theorem1} we know that $\lim\limits_{t_k \to \infty}E[x(t_k)]=\textbf{1}\bar{\pi}_{2}^{T}x(0)$.
Then we analyze $\|e\|=\|\bar{\pi}_{2}^{T}-\frac{\textbf{1}^{T}}{n}\|$. It is clear that $E(x(t_k+\Delta))=W_2E(x(t_k))$ for $h\rightarrow0$, where
\begin{equation}\label{theorem4 newqueation6}
\begin{array}{rl}
W_2=\lim\limits_{h \to 0}W_{1}=e^{-L_{1}\Delta}\alpha+e^{-L_{2}\Delta}\beta+e^{-L_{3}\Delta}\gamma+e^{-L_{4}\Delta}\theta.
\end{array}
\end{equation}
Similar to $W_1$, $W_2$ can
be regarded as the transition matrix of a regular chain. Moreover $W_2=\bar{W}_2+D_2$ and $D_2\textbf{1}=\textbf{0}$.
When $h\rightarrow0$, by using $\lim\limits_{k \to \infty}\bar{W}_{1}^{k}=\frac{\textbf{1}\textbf{1}^{T}}{n}$, we have
$\lim\limits_{k \to \infty}\bar{W}_2=(e^{-L_{4}\Delta})^{k}=\frac{\textbf{1}\textbf{1}^{T}}{n}$. Hence
\begin{equation}\label{theorem4 queation7}
\begin{array}{rl}
\|e\|\leq \sum\limits_{k=0}^\infty\|D_2\bar{W}_2^{k}\|\leq\|D_2\|(1+\bar{\lambda}(\bar{W}_2-\frac{\textbf{1}\textbf{1}^{T}}{n})
+\bar{\lambda}^{2}(\bar{W}_2-\frac{\textbf{1}\textbf{1}^{T}}{n})+\cdots).
\end{array}
\end{equation}
Similar to the above analysis, we get that $0\leq\bar{\lambda}(\bar{W}_2-\frac{\textbf{1}\textbf{1}^{T}}{n})<1$.
Therefore, we have $\|e\|\leq\|D_2\|\frac{1}{1-\bar{\lambda}(\bar{W}_{2}-\frac{\textbf{1}\textbf{1}^{T}}{n})}$.
$\blacksquare$
\begin{remark}
Agent that can't receive information is considered in Scenario I, while
agent that can't send information is considered in Scenario II.
We assume that there exist agents which can
not receive or send information in Scenario III.
For this scenario, similar to Theorem \ref{2016 theorem4}, error bound on consensus value can be calculated.
\end{remark}

\begin{theorem}\label{2016 theorem5}
Assume that the sampling period $\Delta=h<\frac{1}{d_{max}}$ and the expected graph is strongly connected in Scenario I. Then
\begin{equation}\label{theorem5 queation1 }
\begin{array}{rl}
\|\bar{\pi}_{1}^{T}-\frac{\textbf{1}^{T}}{n}\|\leq\dfrac{2c\max\{(\alpha+\beta)^{2},(\beta+\gamma)^{2}\}}{1-\bar{\lambda}(\bar{W}_{1}-\frac{\textbf{1}\textbf{1}^{T}}{n})},
\end{array}
\end{equation}
where $\bar{W}_1=I-hL_{4}$ and $c=h^{2}\max\{\sum\limits_{j=1}^nl_{1j}^{2}, \sum\limits_{j=1}^nl_{2j}^{2}\}$.
\end{theorem}
{\it Proof.}
Due to $h=\Delta$, then we have $W_1=(I-hL_{1})\alpha+(I-hL_{2})\beta+(I-hL_{3})\gamma+(I-hL_{4})\theta$.
Matrix $W_1$ is expressed as $\bar{W}_1+D_1$, where
 \begin{equation}\label{theorem5 queation1}
\begin{array}{rl}
D_1=h\begin{bmatrix}
                                      l_{11}(\alpha+\gamma) & l_{12}(\alpha+\gamma) & \cdots & l_{1n}(\alpha+\gamma)\\
                                      l_{21}(\beta+\gamma) & l_{22}(\beta+\gamma) & \cdots & l_{2n}(\beta+\gamma)\\
                                      0 & 0& \cdots & 0 \\
                                      \vdots & \vdots & \vdots & \vdots \\
                                      0 & 0& 0 & 0 \\
                            \end{bmatrix}.
                             \end{array}
\end{equation}
Similar to the analysis of Theorem \ref{2016 theorem4}, we have $\|e\|=\|\bar{\pi}_{1}^{T}D_1F\|\leq\|D_1F\|=\|D_1\sum\limits_{k=0}^\infty\bar{W}_1^{k}\|$.
To calculate $\|D_1\sum\limits_{k=0}^\infty\bar{W}_1^{k}\|$, we introduce a vector $y(0)$ such that $\|y(0)\|\neq0$ and $y(k+1)=\bar{W}_1^{k}y(0)$.

By utilizing (\ref{theorem5 queation1}), the following equation is obtained
\begin{equation}\label{theorem5 queation2}
\begin{array}{rl}
\|D_1y(k)\|=&h^{2}(\alpha+\gamma)^{2}(l_{11}y_1(k)+l_{12}y_2(k)+\cdots+l_{1n}y_n(k))^{2}\\
&+h^{2}(\beta+\gamma)^{2}(l_{21}y_1(k)+l_{22}y_2(k)+\cdots+l_{2n}y_n(k))^{2}.
\end{array}
\end{equation}
Due to $l_{11}=-\sum\limits_{j=2}^nl_{1j}$ and $l_{22}=-\sum\limits_{j=1, j\neq2}^nl_{2j}$,
\begin{equation}\label{theorem5 queation3}
\begin{array}{rl}
\|D_1y(k)\|=&h^{2}(\alpha+\gamma)^{2}(-l_{12}(y_1(k)-y_2(k))-\cdots-l_{1n}(y_1(k)-y_n(k)))^{2}\\
&+h^{2}(\beta+\gamma)^{2}(-l_{21}(y_2(k)-y_1(k))-\cdots-l_{2n}(y_2(k)-y_n(k)))^{2}.
\end{array}
\end{equation}
Let $\bar{y}(k)=\frac{\textbf{1}^{T}}{n}y(0)$, substituting $\bar{y}(k)$ into equation (\ref{theorem5 queation3}) leads to that
\begin{equation}\label{theorem5 queation4}
\begin{array}{rl}
\|D_1y(k)\|=&h^{2}(\alpha+\gamma)^{2}(l_{11}(y_1(k)-\bar{y}(k))-l_{12}(\bar{y}(k)-y_2(k))-\cdots)^{2}\\
&+h^{2}(\beta+\gamma)^{2}(l_{22}(y_2(k)-\bar{y}(k))-l_{21}(\bar{y}(k)-y_2(k))-\cdots)^{2}\\
\leq& h^{2}(\alpha+\gamma)^{2}\sum\limits_{j=1}^nl_{1j}^{2}\sum\limits_{i=1}^n(y_i(k)-\bar{y}(k))^{2}
+h^{2}(\beta+\gamma)^{2}\sum\limits_{j=1}^nl_{2j}^{2}\sum\limits_{i=1}^n(y_i(k)-\bar{y}(k))^{2}\\
=&h^{2}(\alpha+\gamma)^{2}\sum\limits_{j=1}^nl_{1j}^{2}\|y(k)-\textbf{1}\bar{y}(k)\|
+h^{2}(\beta+\gamma)^{2}\sum\limits_{j=1}^nl_{2j}^{2}\|y(k)-\textbf{1}\bar{y}(k)\|\\
\leq& 2c\max\{(\alpha+\gamma)^{2},(\beta+\gamma)^{2}\}\|y(k)-\textbf{1}\bar{y}(k)\|.
\end{array}
\end{equation}

By using $\|y(k+1)-\textbf{1}\bar{y}(k)\|=\|(W_1-\frac{\textbf{1}\textbf{1}^{T}}{n})y(k)\|$, we get
\begin{equation}\label{theorem5 queation5}
\begin{array}{rl}
\|y(k+1)-\textbf{1}\bar{y}(k)\|\leq\bar{\lambda}^{k}(\bar{W}_{1}-\frac{\textbf{1}\textbf{1}^{T}}{n})\|y(0)\|,
\end{array}
\end{equation}
where $0\leq\bar{\lambda}(\bar{W}_{1}-\frac{\textbf{1}\textbf{1}^{T}}{n})<1$. Hence
\begin{equation}\label{theorem5 queation6}
\begin{array}{rl}
\|D_1y(k+1)\|\leq 2c\max\{(\alpha+\gamma)^{2},(\beta+\gamma)^{2}\}\bar{\lambda}^{k}(\bar{W}_{1}-\frac{\textbf{1}\textbf{1}^{T}}{n})\|y(0)\|.
\end{array}
\end{equation}
Due to $\|D_1\sum\limits_{k=0}^\infty\bar{W}_1^{k}y(0)\|\leq\sum\limits_{k=0}^\infty\|D_1W^{k}y(0)\|$, it holds that
$$\begin{array}{rl}
\sum\limits_{k=0}^\infty\|D_1W^{k}y(0)\|=&\sum\limits_{k=0}^\infty\|D_1y(k)\|\\
\leq&\dfrac{2c\max\{(\alpha+\beta)^{2},(\beta+\gamma)^{2}\}}{1-\bar{\lambda}(\bar{W}_{1}-\frac{\textbf{1}\textbf{1}^{T}}{n})}\|y(0)\|.
\end{array}
$$
Therefore, $\|e\|\leq\|D_1\sum\limits_{k=0}^\infty\bar{W}_1^{k}\|
\leq\dfrac{2c\max\{(\alpha+\beta)^{2},(\beta+\gamma)^{2}\}}{1-\bar{\lambda}(\bar{W}_{1}-\frac{\textbf{1}\textbf{1}^{T}}{n})}$.
$\blacksquare$
\begin{corollary}\label{2016 corollary1}
Assume that the sampling period $\Delta=h<\frac{1}{d_{max}}$ and the expected graph is strongly connected in Scenario II. Then
\begin{equation}\label{corollary1 queation1 }
\begin{array}{rl}
\|\bar{\pi}_{1}^{T}-\frac{\textbf{1}^{T}}{n}\|\leq\dfrac{4\tilde{c}\max\{(\alpha+\beta)^{2},(\beta+\gamma)^{2}\}}{1-\bar{\lambda}(\bar{W}_{1}-\frac{\textbf{1}\textbf{1}^{T}}{n})},
\end{array}
\end{equation}
where $\bar{W}_1=I-hL_{4}$ and $\tilde{c}=h^{2}\max\{\sum\limits_{j=1}^nl_{j1}^{2}, \sum\limits_{j=1}^nl_{j2}^{2}\}$.
\end{corollary}
{\it Proof.}
Due to $h=\Delta$, then we have $W_1=(I-hL_{1})\alpha+(I-hL_{2})\beta+(I-hL_{3})\gamma+(I-hL_{4})\theta$.
Matrix $W_1$ can be expressed as $\bar{W}_1+\tilde{D}_1$, where
 \begin{equation}\label{corollary1 queation1}
\begin{array}{rl}
\tilde{D}_1=h\begin{bmatrix}
                                      -l_{12}(\beta+\gamma) &l_{12}(\beta+\gamma) & 0 &  \cdots & 0\\
                                      l_{21}(\alpha+\gamma) & -l_{21}(\alpha+\gamma) & 0 & \cdots &0\\
                                      l_{31}(\alpha+\gamma) & l_{32}(\beta+\gamma)&  -l_{31}(\alpha+\gamma)
                                      -l_{32}(\beta+\gamma) & \cdots & 0 \\
                                    \vdots& \vdots& \vdots  & \ddots&\vdots\\
                                      l_{n1}(\alpha+\gamma) & l_{n2}(\beta+\gamma) & 0 &  \cdots& -l_{n1}(\alpha+\gamma)-l_{n2}(\beta+\gamma) \\
                             \end{bmatrix}.
                             \end{array}
\end{equation}
By utilizing (\ref{corollary1 queation1}), we have
\begin{equation}\label{corollary queation2}
\begin{array}{rl}
\|\tilde{D}_1y(k)\|=&h^{2}l_{21}^{2}(\alpha+\gamma)^{2}(y_1-y_2)^{2}+h^{2}l_{12}^{2}(\beta+\gamma)^{2}(y_1-y_2)^{2}\\
&+(hl_{31}(\alpha+\gamma)(y_1-y_3)+hl_{32}(\beta+\gamma)(y_2-y_3))^{2}\\
&+\cdots+(hl_{n1}(\alpha+\gamma)(y_1-y_n)+hl_{n2}(\beta+\gamma)(y_2-y_n))^{2}\\
=&(h^{2}l_{21}^{2}(\alpha+\gamma)^{2}+h^{2}l_{12}^{2}(\beta+\gamma)^{2})(y_1-\bar{y}+\bar{y}-y_2)^{2}+\cdots+\\
&(hl_{n1}(\alpha+\gamma)(y_1-\bar{y}+\bar{y}-y_n)+hl_{n2}(\beta+\gamma)(y_2-\bar{y}+\bar{y}-y_n))^{2}\\
\leq& 4h^{2}(\alpha+\gamma)^{2}\sum\limits_{j=1}^nl_{j1}^{2}\sum\limits_{i=1}^n(y_i(k)-\bar{y}(k))^{2}
+4h^{2}(\beta+\gamma)^{2}\sum\limits_{j=1}^nl_{j2}^{2}\sum\limits_{i=1}^n(y_i(k)-\bar{y}(k))^{2}\\
\leq& 4\tilde{c}\max\{(\alpha+\gamma)^{2},(\beta+\gamma)^{2}\}\|y(k)-\textbf{1}\bar{y}(k)\|.
\end{array}
\end{equation}
Similar to the analysis of Theorem \ref{2016 theorem5}, we have
$$\|e\|\leq\|\tilde{D}_1\sum\limits_{k=0}^\infty\bar{W}_1^{k}\|
\leq\dfrac{4\tilde{c}\max\{(\alpha+\beta)^{2},(\beta+\gamma)^{2}\}}{1-\bar{\lambda}(\bar{W}_{1}-\frac{\textbf{1}\textbf{1}^{T}}{n})}.$$
$\blacksquare$


\section{Simulation}
In this section, a simulation is presented
to illustrate the effectiveness of our theoretical results.

\begin{example}\label{2015.example2}
We consider that the communication network is chosen as in Figure \ref{figure-1}.
The interaction topology among agents randomly switches among $G_1$, $G_2$, $G_3$ and $G_4$.
Networks $G_1$, $G_2$, $G_3$ and $G_4$ correspond to the occurrence probabilities
$\alpha=0.3$, $\beta=0.3$, $\gamma=0.2$, $\theta=0.2$, respectively.
By calculation, we can get the sampling period $0<h<0.5$. We choose $h=0.01$, $\Delta=0.1$ and initial value $x(0)=[0.2, 0.8,    0.4,  -1, -2]^{T}$.
Figure \ref{figure-3} depicts the state trajectories of system (\ref{problem statement eq:6}) with random networks.
It can be seen that  all the agents reach consensus.
\begin{figure}
  \centering
  \includegraphics[width=8.5cm]{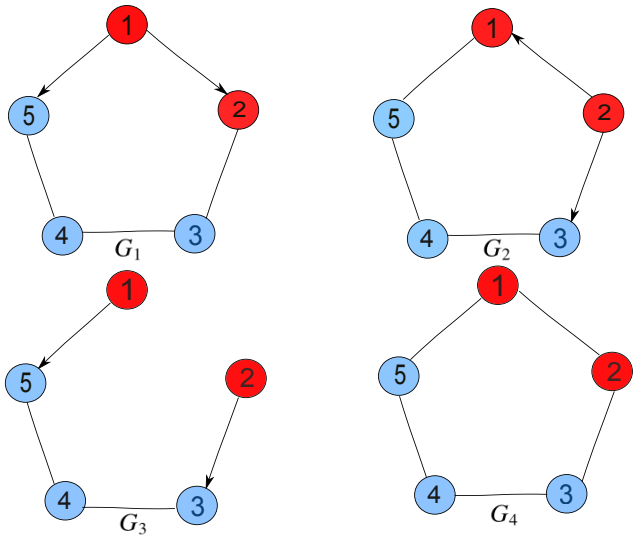}\\
  \caption{Network topologies $G_{1}$, $G_{2}$, $G_{3}$ and $G_{4}$.}\label{figure-1}
\end{figure}
\begin{figure}
  \centering
  \includegraphics[width=10cm]{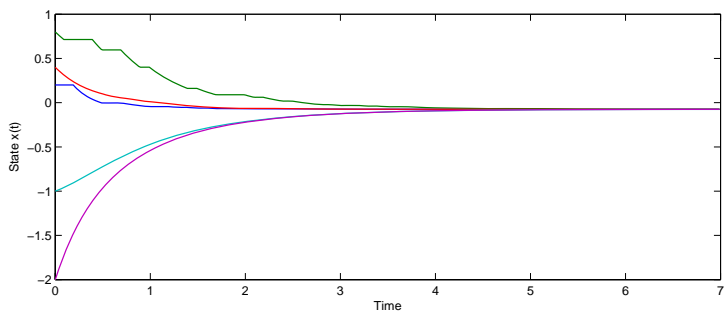}\\
  \caption{State trajectories of system (\ref{problem statement eq:6}) with random networks.}\label{figure-3}
\end{figure}
The original network is denoted by graph $G_4$. The state trajectories of system (\ref{problem statement eq:6}) under network $G_4$ are shown in Figure \ref{figure-2}. It is shown that all the agents reach consensus.
\begin{figure}
  \centering
  \includegraphics[width=10cm]{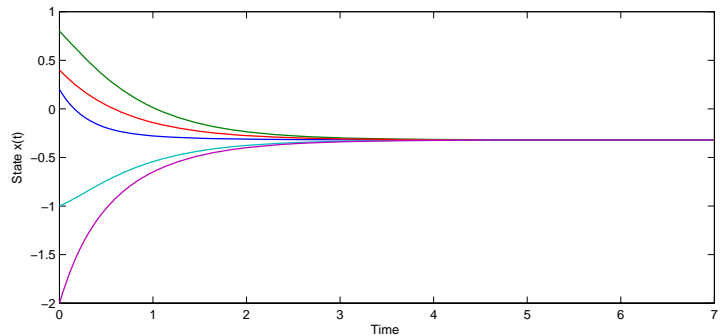}\\
  \caption{State trajectories of system (\ref{problem statement eq:6}) with network $G_4$.}\label{figure-2}
\end{figure}

By calculation, we obtain $\|D_{1}\|\frac{1}{1-\bar{\lambda}(\bar{W}_{1}-\frac{\textbf{1}\textbf{1}^{T}}{n})}=0.0716$.
Therefore, based on Theorem \ref{2016 theorem4}, we can obtain $\|\bar{\pi}^{T}-\frac{\textbf{1}^{T}}{n}\|\leq0.0716$.
It follows from (\ref{Error bound queation2}) that, when ${t_k\rightarrow\infty}$,
$\|E(x(t_k))-\frac{\textbf{1}\textbf{1}^{T}}{n}x(0)\|\leq 2.0918.$
From Figures \ref{figure-3} and \ref{figure-2}, it is easy to verify
that error bound $\|E(x(t_k))-\frac{\textbf{1}\textbf{1}^{T}}{n}x(0)\|$ is less than $2.0918$.
\end{example}
\section {Conclusion}
In this paper, based on the delta operator, a discrete-time multi-agent system is proposed.
It is pointed out that the proposed discrete-time multi-agent system can converge to
the continuous-time multi-agent system as the sampling period tends to zero.
We assume that there exist faulty agents that only send or receive information in the network.
The communication network is described by randomly switching networks.
Under the random networks, it is proved that the consensus in mean (in probability and almost surely) can be achieved when
the expected graph is strong connected.
Furthermore,
the influence of faulty agents on the consensus value is analyzed.
The error bound between consensus values under network with
link failures and original network is presented.
In the future, based on the delta operator, we will consider the formation control and containment control of multi-agent systems, etc.


%

\bibliographystyle{IEEEtran}


\bibliography{BibforControl}

\end{document}